\documentclass[11pt]{article}
\usepackage{amsmath,amsfonts}
\def \beq  {\begin{equation}}
\def \eeq  {\end{equation}}
\def \beqar {\begin{eqnarray}}
\def \eeqar {\end{eqnarray}}
\def\sqr#1#2{{\vcenter{\vbox{\hrule height.#2pt
\hbox{\vrule width.#2pt height#1pt \kern#1pt
\vrule width.#2pt}\hrule height.#2pt}}}}
\def\square{\mathchoice\sqr65\sqr65\sqr{5}3\sqr{5}3}

\def\la {{\langle}}
\def\ra {{\rangle}}
\def\vx {{\vec x}}
\def\vy {{\vec y}}
\def\vk {{\vec k}}
\def\vf {{\varphi}}

\def\Tr {{\rm Tr}}

\def\bD {\bar{D}}

\def\vk {\vec{k}}
\def\vx {{\vec x}}
\def\vz {\vec{z}}
\def\vy{\vec{y}}

\def\vw {\vec{w}}

\def\del {\partial}
\def\bdel{\bar{\partial}}

\def\d {\delta}

\def\bz {{\bar{z}}}

\def\vf {{\varphi}}

\def\half{\textstyle{1\over 2}}

\begin{document}
%%%%%%%%%%%%%%%%%%%%%%%%%%%%%%%%%%%%%%%%%%%%%%%
%\fontfamily{pnb}\fontsize{12pt}{16pt}\selectfont
%\fontfamily{pzc}\fontsize{14pt}{16pt}\selectfont
%\fontfamily{pbk}\fontsize{12pt}{16pt}\selectfont
\fontfamily{cmr}\fontsize{11pt}{14pt}\selectfont
%\fontfamily{phv}\fontshape{ro}\fontsize{11pt}{14pt}\selectfont
%\fontfamily{ptm}\fontseries{m}\fontshape{r}\fontsize{12pt}{16pt}\selectfont
%\fontfamily{pnc}\fontseries{m}\fontshape{r}\fontsize{11pt}{15pt}\selectfont
%\fontfamily{ppl}\fontseries{m}\fontshape{r}\fontsize{11pt}{15pt}\selectfont
%\usefont{T1}{phv}{m}{it}
%%%%%%%%%%%%%%%%%%%%%%%%%%%%%%%%%%%%%%%%%%%%%%%
\def \CMP {{Commun. Math. Phys.}}
\def \PRL {{Phys. Rev. Lett.}}
\def \PL {{Phys. Lett.}}
\def \NPBProc {{Nucl. Phys. B (Proc. Suppl.)}}
\def \NP {{Nucl. Phys.}}
\def \RMP {{Rev. Mod. Phys.}}
\def \JGP {{J. Geom. Phys.}}
\def \CQG {{Class. Quant. Grav.}}
\def \MPL {{Mod. Phys. Lett.}}
\def \IJMP {{ Int. J. Mod. Phys.}}
\def \JHEP {{JHEP}}
\def \PR {{Phys. Rev.}}
\def \JMP {{J. Math. Phys.}}
\def \GRG{{Gen. Rel. Grav.}}
%%%%%%%%%%%%%%%%%%%%%%%%%%%%%%%%%%%%%%%%%%%%%%%
%%%%%%%%%%%%%%%%%%%%%%%%%%%%%%%%%%%%%%%%%%%%%%%
\begin{titlepage}
\null\vspace{-62pt} \pagestyle{empty}
\begin{center}
\rightline{CCNY-HEP-11/6}
\rightline{September 2011}
\vspace{1truein} {\Large\bfseries
The Quantum Effective Action, Wave Functions and}\\
\vskip .1in
{\Large \bfseries Yang-Mills (2+1)}\\
\vskip .1in
{\Large\bfseries ~}\\
%%%%%%%%%%%%%%%%%%%%%%%%%%%%%%%%%%%%%%%%%%%%%%%%\vspace{.6in}
{\large V.P. NAIR}\\
\vskip .2in

{\itshape Physics Department\\
City College of the CUNY\\
New York, NY 10031}\\
\vskip .1in
\begin{tabular}{r l}
E-mail:
&{\fontfamily{cmtt}\fontsize{11pt}{15pt}\selectfont vpn@sci.ccny.cuny.edu}\\
\end{tabular}

\fontfamily{cmr}\fontsize{11pt}{15pt}\selectfont
\vspace{.8in}
%\vspace{1.5in}%\vspace{0.3in}
%%%%%%%%%%%%%%%%%%%%%%%%%%%%%%%%%%%%%%%%%%%%%%%%%%%%%%%%%%%%
\centerline{\large\bf Abstract}
\end{center}
We explore the relationship between the quantum effective action and the ground state (and excited state) wave functions of a field theory. Applied to the Yang-Mills theory in 2+1 dimensions, we find the leading terms of the effective action from the ground state wave function previously obtained in the Hamiltonian formalism by solving the Schr\"odinger equation.

\end{titlepage}

%%%%%%%%%%%%%%%%%%%%%%%%%%%%%%%%%%%%%%%%%%%%%%%%%%%%%%
\pagestyle{plain} \setcounter{page}{2}
\section{Introduction}

This article will be in the nature of continued work on Yang-Mills theories in
2+1 dimensions, along the lines of the Hamiltonian approach initiated a few years ago
\cite{{KN1},{KKN1}, {nair-trento1}}.
Our attempt will be to elucidate a general direct relationship between the quantum effective action and the ground state, and to some extent, the excited state wave functions
of a field theory. The previously obtained ground state wave function for Yang-Mills (2+1) will then be used to identify the leading terms of the effective action for the theory.

We begin with a brief recapitulation of those features of our previous work which are relevant to the present discussion.
The Hamiltonian analysis was done in the $A_0 =0 $ gauge, with the spatial components of the gauge potentials parametrized as
\beq
A_z = {\half} (A_1 + i A_2) = - \del M ~M^{-1}, \hskip .2in A_{\bz} = {\half} (A_1 - i A_2 )
= M^{\dagger -1} \bdel M^\dagger
\label{1}
\eeq
where we use complex coordinates $z = x_1 - i x_2$, $\bz = x_1 + i x_2$.
$M$ is an element of the complexified group; i.e., it is an $SL(N, {\mathbb{C}})$-matrix
if the gauge transformations take values in $SU(N)$.
Wave functions are gauge-invariant and are functions of $H = M^\dagger M$, with the inner product
\beq
\la 1\vert 2\ra = \int d\mu (H) \exp [2~c_A~S_{wzw}(H)]~ \Psi_1^* \Psi_2
\label{2}
\eeq
where $S_{wzw}$ is the Wess-Zumino-Witten action given by
\beq
S_{wzw} (H) = {1 \over {2 \pi}} \int \Tr (\partial H ~\bdel
H^{-1}) +{i \over {12 \pi}} \int \epsilon ^{\mu \nu \alpha} \Tr (
H^{-1}
\partial _{\mu} H~ H^{-1}
\partial _{\nu}H ~H^{-1} \partial _{\alpha}H)
\label{3}
\eeq
In equation (\ref{2}), $d\mu (H)$ is the Haar measure for the gauge-invariant variable
$H$ which takes values in $SL(N, {\mathbb{C}})/SU(N)$. Further, $c_A$ is the value of the quadratic Casimir operator for the adjoint representation;
it is equal to $N$ for $SU(N)$.
The Hamiltonian and other observables can be taken to be functions of the current $J$ of the WZW action, namely, of
\beq
J = {2 \over e} \, \del H ~ H^{-1}
\label{4}
\eeq
(This is not exactly the current as conventionally defined, we have multiplied by some constant factors to simplify some formulae later.) Explicitly,
${\cal H} = {\cal H}_0 + {\cal H}_1$, where
\beqar
{\cal H}_0 &=& m  \int_z J_a (\vz) {\d \over {\d J_a (\vz)}} + {2\over \pi} \int _{z,w} 
 {1\over (z-w)^2} {\d \over {\d J_a (\vw)}} {\d \over {\d
J_a (\vz)}}\nonumber\\
&&\hskip .4in + {1\over 2} \int_x :\bdel J^a(x) ~\bdel J^a(x):\label{5}\\
{\cal H}_1& =&  i ~{e}~ f_{abc} \int_{z,w}  {J^c(\vw) \over \pi (z-w)}~ {\d \over {\d J_a (\vw)}} {\d \over {\d
J_b (\vz)}} \nonumber
\eeqar
where $m = e^2 c_A /2\pi$.

Our basic strategy was to solve the Schr\"odinger equation keeping all terms in ${\cal H}_0$
at the lowest  order and treating ${\cal H}_1$ as a perturbation. Since $m = e^2 c_A/2\pi$,
in ordinary perturbation theory, one would expand in powers of $m$ as well. So our expansion corresponds to a partially resummed version.
Formally, we keep $m$ and $e$ as independent parameters in keeping track of different orders, only setting $m = e^2 c_A/2\pi$ at the end.
The lowest order computation of the wave function in this scheme was given in \cite{KKN2} and gave the string tension as $\sigma_R = e^4 c_A c_R /4 \pi$. More recently, we calculated corrections to this formula, taking the expansion to the next higher order (which still involves an infinity of correction terms) and found that these were small, of the order of 
$-0.03\%$ to $-2.8\%$ \cite{KNY}.

We shall also recall briefly a short argument from \cite{KNrobust} on the nature of the wave function.
For this, absorb the factor $e^{2 c_A S_{wzw}}$ in (\ref{2}) into the definition of the wave function by writing $\Psi = e^{- c_A S_{wzw}} \, \Phi$. The Hamiltonian acting on $\Phi$ is given by
${\cal H} \rightarrow e^{- c_A S_{wzw}}\, {\cal H}\, e^{- c_A S_{wzw}}$.
We now expand $H$ as  $H = \exp( t_a \vf^a ) \, \approx  1 + t_a \vf^a + \cdots$; this ``small $\vf$" expansion is suitable for a (resummed) perturbation theory. The Hamiltonian is then
\beq
{\cal H}= {1\over 2}\int \left[ -{\delta^2 \over \delta \phi^2} +\phi (-\nabla^2 +m^2)\phi
+\cdots\right] \label{6}
\eeq
where $\phi _a (\vk) = \sqrt {{c_A k \bar{k} }/ (2 \pi m)}~~ \vf _a (\vk)$. This is the Hamiltonian for a field of mass $m$ and gives the vacuum wave function 
\beq
\Phi_0 \approx \exp \left[ - {1\over 2} \int \phi^a \sqrt{ m^2 - \nabla^2} ~\phi^a \right]
\label{7}
\eeq
Transforming back to the $\Psi$'s, we find
\beq
\Psi_0 \approx \exp \left[ - {c_A \over \pi m} \int (\bdel \del \vf^a) \left[{1\over  \sqrt{- \nabla^2 + m^2}~ + m}\right]
(\bdel \del \vf^a) + \cdots \right] \label{8}
\eeq
Now comes the key argument:
On general grounds, see \cite{{KKN1},{KNrobust}}, the  full wave function must be a functional of the current $J$. So we can ask: Is there a functional of the current $J$ which reduces to
(\ref{8}) in the small $\vf$ approximation, when $J^a \approx (2/ e) \del \vf ^a+{\cal O}(\vf^2)$?
The only form consistent
with this is
\beq
\Psi_0  =  \exp \left[ - {2\pi^2 \over e^2 c_A^2} \int \bdel J^a (x) \left[{1\over  \sqrt{- \nabla^2 + m^2}~ + m}\right]_{x,y} ~\bdel J^a (y) ~+\cdots\right]
\label{9}
\eeq
This is, of course, the wave function we found by directly solving the Schr\"odinger equation,
${\cal H}_0 \, \Psi_0 \approx 0$. Notice also that we may write this wave function as
\beq
\Psi_0  =  \exp \left[ - {2\pi^2 \over e^2 c_A^2} \int \bdel J^a (x) \left[{\sqrt{k^2 + m^2}~  { {- m}} \over k^2}\right]_{x,y} ~\bdel J^a (y) ~+\cdots\right]
\label{10}
\eeq
In the integral kernel in the exponent, the term $\sqrt{k^2 + m^2} \, / k^2 $ is due to the fact that we have a mass for the fields $\phi$, while the second part $- m /k^2$ is from transforming using $e^{c_A S_{wzw}}$ from the measure. This argument for $\Psi_0$ thus emphasizes the role of the measure in both generating a mass $m$ and in providing the crucial $-m/k^2$ term. The latter is important in obtaining the low momentum limit 
\beq
{ \sqrt{k^2 + m^2}\, - m \over k^2} \approx {1\over 2 m}
\label{11}
\eeq
so that the exponent in $\Psi_0^* \Psi_0$ is the two-dimensional Yang-Mills action,
$\int \bdel J \bdel J \sim \int F^2 / 4 g^2$, $g^2 = m e^2$.
This was, in turn, the key to obtaining the formula for the string tension.

We can now phrase the basic question we address in this paper: Can we find an effective three-dimensional action which will give this wave function including the crucial $- m /k^2$ term in the kernel?
We are focusing on terms to the quadratic order in the currents or gauge potentials, so that it is useful to rewrite $\Psi_0$ as
\beq
\Psi_0 \approx  \exp \left[ - {1\over 2} \int A^{aT}_i (x) \left[{\sqrt{k^2 + m^2}~  {{- m}} }\right]_{x,y} ~A^{aT}_i (y) ~+\cdots\right]
\label{12}
\eeq
where we use the transverse component of $A_i^a$ as the gauge-invariant variable; this is an adequate representation for our argument to the quadratic order.

\section{The effective action and wave functions}

\noindent{\underline{Ground state wave function}}

Starting from the Yang-Mills action, we can construct the Hamiltonian operator and 
solve the Schr\"odinger equation to find the ground state wave function. This is the path we have followed in previous work. As for the effective action, it will include a gauge-invariant mass term for the fields,
which must be nonlocal, including nonlocality in time.
The Hamiltonian set up is thus nontrivial.
Of course, the effective action has the quantum 
dynamics built in, so we should not quantize it. 
Nevertheless, being nonlocal, even a
classical Hamiltonian formulation is not simple.
We will need a more direct way to connect the quantum effective action and wave functions.
This can be done as follows. We will use a scalar field to illustrate this basic connection.
First of all, by using a complete set of energy states $\vert \alpha \ra$, we can write
\beqar
\la \vf \vert e^{-  \beta \, {\cal H}} \vert \vf' \ra &=&
\sum_\alpha \la \vf \vert \alpha \ra \, \la \alpha \vert \vf'\ra\, e^{-\beta E_\alpha}
=
\sum_\alpha \Psi_\alpha (\vf) \, \Psi^*_\alpha (\vf') \, e^{- \beta E_\alpha}\nonumber\\
&\rightarrow& \Psi_0 (\vf ) \, \Psi^*_0 (\vf') \, e^{-\beta E_0},\hskip .3in {\rm as}~~ {\beta \rightarrow \infty}
\label{13}
\eeqar
So we can extract $\Psi_0(\vf )$ by calculating this matrix element with fixed boundary values of the field at the Euclidean time-boundaries, $\tau =0, \, \beta$.
The second step is to write this matrix element as a functional integral,
\beq
\la \vf \vert e^{-  \beta \, {\cal H}} \vert \vf' \ra  = \int [d\phi]\, e^{- S(\phi)} 
= \int [d\eta] \, e^{- S ( \chi + \eta )}
\label{14}
\eeq
The boundary conditions on $\chi (\tau, \vx )$ and $\eta (\tau, \vx )$ are
\begin{align}
\chi (0 , \vx ) &= \vf' (\vx ), \hskip .3in \chi (\beta, \vx ) = \vf (\vx )\nonumber\\
\eta (0, \vx ) &= \eta (\beta, \vx )  = 0
\label{15}
\end{align}
$\chi (\tau, \vx )$ is a fixed field configuration with the boundary values specified; it contains no additional degree of freedom to be integrated in (\ref{14}). Since $\chi$ gives the requisite boundary behavior, $\eta$ must vanish at both $\tau =0$ and $\tau = \beta$. Thus, in carrying out the $\eta$-integration in (\ref{14}), we must use Dirichlet conditions in $\tau$ for the $\eta$-propagator. However, rather than explicitly carrying out the $\eta$-integration, we may note that the quantum effective action $\Gamma [\chi ]$ is defined, {\it for arbitrary} $\chi$,
by
\beq
e^{- \Gamma (\chi)} = \int [d\eta] \, \exp\left[ - S(\chi + \eta ) + \int {\delta \Gamma \over \delta \chi}
\,\eta \right]
\label{16}
\eeq
From this equation, we see that, if we choose $\chi$ as a solution of $\delta \Gamma /\delta \chi \,
= 0$, with the boundary behavior $\chi \rightarrow \vf'$ at $\tau =0$ and
$\chi \rightarrow \vf $ at $\tau = \beta$, and with $\eta$ going to zero at both ends,
then
\beqar
e^{- \Gamma } &=& \int [d\eta]\, e^{- S (\chi +\eta )} = \la \vf\vert e^{-\beta {\cal H}} \vert \vf'\ra\nonumber\\
&\rightarrow& \Psi_0 (\vf) \Psi^*_0 (\vf') \, e^{-\beta E_0}, \hskip .2in {\rm as}~ \beta \rightarrow \infty
\label{17}
\eeqar
In other words, if we solve the equation
\beq
{\delta \Gamma \over \delta \chi} = 0
\label{18}
\eeq
for $\chi$, subject to the boundary conditions (\ref{15}), and substitute this back in $\Gamma (\chi )$,
then $e^{-\Gamma(\chi)}$ which is now a functional of $\vf'$, $\vf$, will give
 $\Psi_0 (\vf )$ as $\beta$ becomes large. This relates $\Psi_0 (\vf)$ and $\Gamma(\chi)$ directly.
 
 We want to emphasize that, depending on the boundary conditions used for the $\eta$'s
 in carrying out the functional integration in (\ref{16}), there are different
 $\Gamma$'s we can define. For our purpose,
  to get agreement between (\ref{14}) and $\Gamma$ as defined by
 (\ref{16}), the $\eta$'s in the functional integral in (\ref{16}) must vanish at
 $\tau = 0, \, \beta$. As a result, the Green's functions which may occur in $\Gamma$ obey Dirichlet boundary conditions for the Euclidean time-direction.
 We can see this more explicitly by considering an example, say, a $\phi^4$-theory
 with the action
 \beq
 S = \int \left[ {1\over 2} \left( {\dot \phi}^2 + (\nabla \phi )^2 + \mu^2 \phi^2 \right) 
 + \lambda \, \phi^4 \right]
 \label{18a}
 \eeq 
 The effective action $\Gamma$ can be examined in a loop expansion
 $\Gamma = S + \hbar \, \Gamma^{(1)} + \hbar^2 \, \Gamma^{(2)} + \cdots$. Using this in (\ref{16}) we find
 \beqar
 \Gamma^{(1)} (\chi ) &=& - \log \left[ \int [d\eta ] \, e^{- {\half} \int \eta (x) \, M(x,y) \, \eta (y) }\right]
 =  {1\over 2} \log \det M\nonumber\\
 M(x, y) &=& \left[ {\delta^2 S \over \delta \phi(x) \delta \phi (y) }\right]_{x,y}
 = \left[ ( - \square + \mu^2 ) + 12 \, \lambda \, \chi^2(x) \right]\, \delta (x-y)
 \label{18b}
 \eeqar
 The determinant must be evaluated using eigenfunctions which vanish at
 $\tau = 0, \, \beta$, since $\eta$'s obey this condition.
 For the contributions from the ${\cal O}(\eta^3 )$ terms which give the higher loop terms, we will need the
 inverse of $M$ which can be expanded as
 \beq
 M^{-1}(x,y) = G(x, y, \mu) - \int_z G(x, z, \mu) \, [12 \lambda \chi^2(z) ]\, G(z, y, \mu) + \cdots
 \label{18c}
 \eeq
 where $G(x, y, \mu) = ( - \square + \mu^2 )^{-1}$. This Green's function must also vanish at
 $\tau = 0, \, \beta$.
 We see from this procedure that all the Green's functions appearing in $\Gamma (\chi)$ so evaluated will obey Dirichlet conditions at $\tau =0, \, \beta$. Basically this means that the expression for $\Gamma$ will be identical to the usual one, except that the 
 Feynman propagators (or their Euclidean versions)
 will be replaced by their Dirichlet versions.
 
 In practice, the evaluation of $\Gamma$ on its critical point can be simplified a bit further, at least for the case of interest to us in what follows.
 Let $W$ denote $\Gamma$ evaluated on the solution $\chi_*$ of (\ref{18}),
 subject to the boundary values (\ref{15}).
If we vary the boundary value $\vf$ of $\chi$ and also change $\beta$ slightly, the resulting variation of $\Gamma$ or $W$ can be written in the form
\beq
\delta W = \delta \Gamma[\chi_*] = \int d^2x~ \Pi \, \delta \vf ~+~ {\cal H}_E \, \delta \beta
\label{19}
\eeq
This defines $\Pi$ (which may depend on the time-derivatives of $\vf$) and also the Euclidean Hamiltonian ${\cal H}_E$, which is generally not positive semi-definite.
Since we are evaluating $\Gamma$ on the solution of (\ref{18}), the terms involving 3d-volume integrals are zero.

Generally, ${\cal H}_E$ will give the zero-point energy, but for a relativistically invariant vacuum, we know that the zero-point energy  must be zero.
Therefore, we can impose ${\cal H}_E = 0$. Further, $\Pi$ may be taken as $\delta W / \delta \vf$.
Thus we can find $W$ by solving the equations
\beq
{\cal H} _E = 0, \hskip .3in \Pi = {\delta W \over \delta \vf}
\label{20}
\eeq
The ground state wave function is then given by $ \Psi_0 = e^{-W}$. Needless to say, this
is a Euclidean version of the usual Hamilton-Jacobi approach.

It is useful to work this out in a simple example such as the $\phi^4$-theory.
The effective action $\Gamma$ for this theory is of the form
\beq
\Gamma = \int \, {1\over 2} \left( {\dot \chi}^2 + (\nabla \chi )^2 + \mu^2 \chi^2\right)
+ \int \, V( x_1, x_2, x_3, x_4) \, \chi(x_1) \chi(x_2) \chi (x_3) \chi(x_4) + \cdots
\label{20a}
\eeq
where $V(x_1, \cdots, x_4)$ and higher point terms are nonlocal vertices.
The variation at the time-slice $\tau = \beta$ gives
\beq
\delta \Gamma =  \int {\dot \chi} \, \delta \chi + {\cal H}_E \, \delta \beta~
=  \sum_k {\dot c}_k \, d c_k  + {\cal H}_E \, \delta \beta
\label{20b}
\eeq
where we introduced a mode expansion $\chi = \sum_k c_k \, u_k (x)$ in terms of the eigenmodes of $\nabla^2$ and
\beqar
{\cal H}_E \!\!\!&=& \!\!\! - {1\over 2} \int  {\dot \chi}^2 +  \left[ {1\over 2}\int
\left( (\nabla \chi )^2 + \mu^2 \chi^2 \right) + \int V( x_1, x_2, x_3, x_4) \, \chi(x_1) \chi(x_2) \chi (x_3) \chi(x_4) + \cdots\right]\nonumber\\
&=&\!\!\! - {1\over 2} \sum_k {\dot c}_k^2 + \left[ {1\over 2} \sum_k \omega_k^2\, c_k^2 
+ \sum_{\{k_i\}} V(k_1, k_2, k_3, k_4) c_{k_1} c_{k_2} c_{k_3} c_{k_4} + \cdots\right]
\label{20c}
\eeqar
with $\omega_k^2 = k^2 + \mu^2$.
The Hamilton-Jacobi equation thus reduces to
\beq
{1\over 2} \sum_k \left( {\del W \over \del c_k}\right)^2 =
 \left[ {1\over 2} \sum_k \omega_k^2\, c_k^2 
+ \sum_{\{k_i\}} V(k_1, k_2, k_3, k_4) c_{k_1} c_{k_2} c_{k_3} c_{k_4} + \cdots\right]
\label{20d}
\eeq
By taking an ansatz for $W$ as a power series in the $c_k$'s and treating $V$ perturbatively, this is easily solved as
\beq
W = {1\over 2} \sum_k \omega_k \, c_k^2 +
\sum_{\{k_i\}}  {V(k_1, k_2, k_3, k_4)\over
\omega_1 + \omega_2 + \omega_3 + \omega_4} \, c_{k_1} c_{k_2} c_{k_3} c_{k_4} + \cdots
\label{20e}
\eeq

\noindent{\underline{Excited states}}

The ground state wave function also contains some information about the excited states.
So, once we have obtained $\Psi_0$ (or $W$) from the quantum effective action
$\Gamma$, we can set up Schr\"odinger equations involving excited states as follows.
We illustrate this by considering a scalar field theory again, taking the action as
\beq
 S_M = \int \left[ {1\over 2} {\dot \phi}^2 ~-~ U(\phi)\right]
 \label{20f}
 \eeq 
where $U(\phi)$ contains the spatial derivative terms and interaction terms (which could be something more involved than $\phi^4$). The subscript $M$ on $S$ is to emphasize that we are in Minkowski space now.
Given such an action, we can, in principle, determine $\Gamma$ and eventually $\Psi_0$ as outlined above.
Now consider a slightly modified action
\beq
{\widetilde S}_M = S_M ~+~ \int \xi (\vx)\, {\dot \phi}
\label{20g}
\eeq
where $\xi$ is an external source taken to be independent of time, so that the last term is actually a total derivative.
In carrying out the quantization of this action, we find
\beqar
{\dot \phi} &=& -i {\delta \over \delta \phi} ~-~ \xi\nonumber\\
{\widetilde {\cal H}}_M &=& \int \left[ {1\over 2} {\dot\phi}^2 + U(\phi)\right]
= \int \left[ - {1\over 2} {\delta^2 \over \delta \phi ^2 }+ U(\phi)\right] ~+~ i \int \xi {\delta \over \delta \phi} + {1\over 2} \int \xi^2\nonumber\\
&=& {\cal H}_M ~+~  i \int \xi {\delta \over \delta \phi} + {1\over 2} \int \xi^2
\label{20h}
\eeqar
Since we have added an external source, we do not have an argument for Lorentz invariance and hence it is not {\it a priori} obvious that the ground state energy is zero. Let ${\widetilde \Psi}_0$ be the new ground state wave function and $E_0 (\xi)$ (which may depend on $\xi$) be the new ground state energy. We can then write
\beq
\left[ {\cal H}_M ~+~  i \int \xi {\delta \over \delta \phi} + {1\over 2} \int \xi^2\right]
\, {\widetilde \Psi}_0 = E_0 (\xi) \, {\widetilde \Psi}_0
\label{20j}
\eeq

We now consider the matrix element of $\exp (- \beta {\widetilde{\cal H}}_M)$ and taking $\xi$ to be a small enough perturbation that there is still a ground state, we can write
\beq
\la \vf \vert e^{-  \beta \, {\widetilde{\cal H}}_M } \vert \vf' \ra 
\rightarrow {\widetilde\Psi}_0 (\vf ) \, {\widetilde\Psi}^*_0 (\vf') \, e^{-\beta E_0(\xi)},\hskip .3in {\rm as}~~ {\beta \rightarrow \infty}
\label{20k}
\eeq
Now, once again, we write the left hand side as a functional integral,
\beqar
\la \vf \vert e^{-  \beta \, {\widetilde{\cal H}}_M } \vert \vf' \ra 
&=& \int [d\phi]\, \exp \left( - S_E (\phi) + i \int \xi {\dot \phi} \right)\nonumber\\
&=& e^{i \int \xi \vf } \, \int [d\phi]\, e^{- S_E (\phi)} \, \, e^{-i\int \xi \vf'}\nonumber\\
&=& e^{i \int \xi \vf } \, \la \vf \vert e^{-  \beta \, {{\cal H}}_M } \vert \vf' \ra \,\,
 e^{-i\int \xi \vf'}\nonumber\\
 &\rightarrow&  e^{-\beta E_0 (\xi =0)}\,\, \left[e^{i \int \xi \vf } \,\Psi_0 (\vf) \right]\, \, 
 \left[e^{ i \int \xi \vf' }\, \Psi_0 (\vf ') \right]^*
 \label{20m}
\eeqar
where it is implicit in the functional integrals in the first and second lines of this equation that the boundary conditions are $\phi = \vf$ at $\tau = \beta$ and $\phi =\vf'$ at $\tau =0$.
Comparing (\ref{20k}) and (\ref{20m}), we see that we can still take $E_0 (\xi )$ to be zero,
since $E_0 (\xi =0 )$ is zero by the Lorentz invariance argument; further, 
\beq
{\widetilde \Psi}_0 (\vf ) = \exp \left(i \int \xi\, \vf \right)\,\, \Psi_0 (\vf )
\label{20n}
\eeq
The Schr\"odinger equation for ${\widetilde \Psi}_0$, namely, equation (\ref{20j}), then becomes
\beq
{\cal H}_M  \left[ e^{i \int \xi\, \vf} \, \Psi_0 \right]
= \left[ {1\over 2} \int \xi^2 + i \int \xi\, {\delta W \over \delta \vf} \right] \, e^{i \int \xi\, \vf} \, 
\, \Psi_0 \label{20p}
\eeq
where we have used the expression $\Psi_0 = \exp (- W )$.
The expansion of this equation in powers of $\xi$ will give a set of equations which correspond to the Schr\"odinger equation for excited states. The basic ingredient which went into this equation is that the action is quadratic in the time-derivatives. (Otherwise we will get additional terms
involving $\xi$'s.) The ground state wave function determines the nature of various terms in this equation via
the function $W$.

It is instructive to see how the Schr\"odinger equation
(\ref{20p}) works out in a simple case, say, for the theory given by (\ref{18a}).
In this case, $W$ is given by
(\ref{20e}), which we write as $W =  {\half} \sum_k \omega_k \, c_k^2 \, +\, W_1$. The $\xi$-independent term of (\ref{20p}) gives just the expected result ${\cal H}_M \, \Psi_0 = 0$.
The term linear in $\xi$ gives
\beq
{\cal H}_M \, \left[ c_k \, \Psi_0 \right] = \omega_k \, \left[ c_k \, \Psi_0 \right] 
~+~ {\del W_1 \over \del c_k } \, \Psi_0
\label{20q}
\eeq
If interactions are ignored, we get the expected one-particle result.
The term involving $W_1$ shows that this state mixes with the higher states.
Likewise, the terms quadratic in $\xi$ gives
\beq
{\cal H}_M \, \left[ c_k \, c_l \, \Psi_0 \right] = \left[ ( \omega_k + \omega_l ) \, c_k \, c_l\,
- \delta_{kl} + c_k {\del W_1 \over \del c_l} + c_l {\del W_1 \over \del c_k} \right] \, \Psi_0
\label{20r}
\eeq
The state $c_k \, c_l \, \Psi_0$ is not orthogonal to the ground state. Let
\beq
\la c_k \, c_l \ra \equiv \int \Psi_0^* \, c_k \, c_l \, \Psi_0
= \int [dc]\, \exp\left( - \sum_k \omega_k c_k^2 - 2 \, W_1\right) \, c_k \, c_l
\label{20s}
\eeq
We can rewrite equation (\ref{20r}) as
\beqar
{\cal H}_M \, \left[ \left( c_k \, c_l  - \la c_k c_l \ra \right) \Psi_0 \right] &=&
(\omega_k + \omega_l ) \, \left[ \left( c_k \, c_l  - \la c_k c_l \ra \right) \Psi_0 \right] \nonumber\\
&&\hskip .1in + \left[ ( \omega_k + \omega_l ) \la c_k c_l\ra  - \delta_{kl} +  c_k {\del W_1 \over \del c_l} + c_l {\del W_1 \over \del c_k} \right]\, \Psi_0
\label{20t}
\eeqar
In the absence of interactions $\la c_k c_l \ra = \delta_{kl} /2 \omega_k$, so the second line is zero and the equation correctly gives the two-particle eigenstate. With interactions, the second line describes possible mixing with other higher states.

It is clear that the process can be continued to obtain equations for higher states.
The action of ${\cal H}_M$ on a given state has other orthogonal states on the right hand side.
So while we do not have a diagonal form for ${\cal H}_M$,  the point is that all matrix elements
of ${\cal H}_M$ are determined by the ground state wave function.
Notice that the expectation values needed for orthogonalization are calculated with the full
ground state wave function.
This does have implications for the simplification of the higher terms in (\ref{20t}).
For example, we can have a term with four $c$'s, such as
$K(k_1, k_2, k_3, k_4) \, c_{k_1} \, c_{k_2}\, c_{k_3}\, c_{k_4}$ on the right hand side arising from $W_1$, where $K(k_1, k_2, k_3, k_4)$ is the appropriate kernel.
This means that the two-particle equation mixes with the four-particle states. In a truncation to the two-particle level, we can approximate the product of the $c$'s as
\beq
c_{k_1} \, c_{k_2}\, c_{k_3}\, c_{k_4} \approx c_{k_1} \, c_{k_2}\, \la c_{k_3}\, c_{k_4}\ra 
+ {\rm permutations}
\label{20u}
\eeq
(This is very much in the spirit of an operator product expansion for the product of the $c$'s.)
The result is then a two-particle equation with the constituent particles interacting via a potential
\beq
V \approx \int_{k_3, k_4}\, K(k_1, k_2, k_3, k_4) \,  \la c_{k_3}\, c_{k_4}\ra ~
+{\rm permutations}
\label{20v}
\eeq
The expectation value $\la c_{k_3} \, c_{k_4}\ra$ is calculated with the full ground state wave function and it determines the potential involved in the construction of the higher excited states.

\noindent{\underline{Summary}}

We now briefly recapitulate the results of this section.
\begin{itemize}
\item To find the ground state wave function:
\begin{itemize}
\item We begin with the Euclidean quantum effective action $\Gamma$ calculated with
Dirichlet boundary conditions in the time-direction
\item Find the solution $\chi^*$  of $(\delta \Gamma /\delta \chi) \, =0$ with the boundary conditions
$\chi (0, \vx ) = \vf'(\vx)$, $\chi (\beta , \vx ) = \vf (\vx)$.
\item $\exp( - \Gamma ( \chi^*) )$ then gives the ground state wave function, up to normalization, as $\beta$ becomes large.
\item Alternatively, we can solve the Euclidean Hamilton-Jacobi equation
${\cal H}_E = 0$, $\Pi = \delta W /\delta \vf$, where
${\cal H}_E$ and $\Pi$ are defined by (\ref{19}). $e^{-W}$ then gives the ground state wave function.
\end{itemize}
\item For the excited states:
\begin{itemize}
\item Once $\Psi_0$ is obtained, we construct the Schr\"odinger equation
(\ref{20p}). Expansion in powers of $\xi$ will give a series of equations.
\item These are not yet eigenstates of the Hamiltonian, a rediagonalization is, in general, needed.
This equation basically gives us the matrix elements of the Hamiltonian in a chosen basis.
It is $W$ which determines the nature of this equation and hence some nonperturbative information can be built in via this function if we have a way of obtaining it nonperturbatively.
\item The procedure can be generalized to obtain excited states which are given by composite operators, rather than powers of $\vf$, acting on $\Psi_0$ and to cases where the time-derivative in the Hamiltonian is not a simple quadratic form.
\end{itemize}
\end{itemize}

\section{The effective action for Yang-Mills (2+1)}

We are now in a position to state the main result of this paper. The leading terms of the quantum effective
action for 3-dimensional Yang-Mills theory are given by
\beq
\Gamma = \int {1\over 4} F^a_{\mu\nu} F^a_{\mu\nu} + S_{m}(A) 
+ (\sigma^\mu D_\mu \Phi_A)^{a\dagger} (\sigma^\nu D_\nu \Phi_A)^{a} ~+~ \cdots
\label{21}
\eeq
where $S_m(A)$ is a gauge-invariant nonlocal mass term for the gauge field.
The particular choice of this mass term is not important at this stage.
We will discuss this later.
$\Phi^a_A$, $a = 1, 2, \cdots, (N^2 -1)$,  $A= 1, 2$, is a complex field transforming according to the adjoint representation of $SU(N)$, and transforming as a 2-component spinor under the Lorentz group.
$\sigma^\mu$, $\mu = 1, 2, 3$, are the Pauli matrices and $D_\mu$ denotes the gauge-covariant derivative. A complex spinor field with a quadratic derivative term in the action is unusual, but it is not to be considered as an observable field. It is to be viewed simply as a method of capturing the physics of the wave function (\ref{11}) or (\ref{12}). The action has an additional $U(1)$ symmetry $\Phi \rightarrow e^{i \theta} \, \Phi$, which the original Yang-Mills theory does not have. We will eliminate this unwanted symmetry by requiring that all physical operators must have equal numbers of 
$\Phi$'s and $\Phi^*$'s.

We will first show how this action leads to the wave function (\ref{12}), before discussing further properties.
The equations of motion corresponding to (\ref{21}) are
\beqar
- (D_\mu F_{\mu\nu} )^a ~+~ {\delta S_m \over \delta A_\nu^a} &=& e \left( (D_\mu \Phi)^\dagger
T^a \Phi - \Phi^\dagger T^a D_\mu \Phi \right)\label{22}\\
D_\mu \left( \sigma^\mu \, \sigma^\nu \, D_\nu \Phi \right) & =& 0\label{23}
\eeqar
where $\{ T^a\}$ are a basis of the Lie algebra generators in the adjoint representation.
In the first equation, we will keep the mass term at the lowest order, but treat the effect of the current due to $\Phi$ (the right hand side of (\ref{22})) in a perturbation expansion.
We will solve the second equation as it is.  This expansion scheme is thus similar to what we did in the Hamiltonian approach in \cite{{KKN2}, {KNY}}.
This means that we can treat the Yang-Mills part and the $\Phi$-dependent terms of $\Gamma$ in
(\ref{21}) separately to the lowest order.
Also we may just retain the terms linear in $A$ on the left hand side of (\ref{22}) (or terms quadratic in $A$ at the level of $\Gamma$) to the same order.
The quadratic term in $S_m (A)$, for any choice of the mass term, has the same form, namely,
$\sim A^{T2}$. Writing $A^T_\mu = A_\mu - \int_y \, \del_\mu G(x, y) \,\del \!\cdot \!\!A (y)$, we see that it is invariant under the (Abelian) gauge transformation $A_\mu \rightarrow A_\mu + \del_\mu \theta$, provided $G(x,y)$ obeys Dirichlet conditions and $\theta$ vanishes at $\tau =0, \beta$. In this case, we can write
\beq
S_m (A) = {m^2 \over 2} \int A^{T2} ~+\cdots = {m^2 \over 2} \int \left[ A^2 - \del\!\cdot \!\!A (x) \, G(x,y) \, \del\!\cdot \!\!A (y) + \cdots \right]
\label{24}
\eeq
For the Yang-Mills part of the action, we then find
\beq
\delta W_{YM} = \int d^2x~ F^T_{0i} \delta A^T_i ~+~ \int d^2x~ {1\over 2} \left[ - F_{0i}^2 + A^T_i ( k^2 + m^2) A^T_i \right] \, \delta \beta
\label{25}
\eeq
Setting ${\cal H}_E$ to zero, we find
\beq
W_{YM} =  {1\over 2} \int d^2x~ A^T_i \sqrt{k^2 + m^2}\, A^T_i ~+\cdots
\label{26}
\eeq
This is entirely as expected. In the $A_0 =0$ gauge, for the $\Phi$-dependent terms, we find
\beqar
\delta W &=& \int \left[ \delta \phi_1^\dagger ( {\dot \phi}_1 + 2\, {\bar D} \phi_2 ) + \delta \phi_2^\dagger
({\dot \phi}_2 - 2 \, D \phi_1) ~+ ~c.c.\right] ~+~ {\cal H}_E \, \delta \beta\nonumber\\
{\cal H}_E &=&\int \left[ 4 ({\bar D} \phi_2)^\dagger ({\bar D} \phi_2) + 4 ({\bar D} \phi_1)^\dagger
({\bar D} \phi_1) - {\dot \phi}_1 {\dot \phi}_1 - {\dot \phi}_2 {\dot \phi}_2 \right]
\label{27}
\eeqar
Solving ${\cal H}_E = 0$, we find
\beqar
W_\Phi &=& \Phi^\dagger \, K \, \Phi\nonumber\\
K&=& 4\, \left[\begin{matrix}
~~0 &~ {\bar D} \\
- D & ~0\\
\end{matrix}\right], \hskip .3in \Phi^a = \left( \begin{matrix}
\phi^a_1\\
\phi^a_2\\
\end{matrix}
\right)
\label{28}
\eeqar
As mentioned above, the field $\Phi^a_A$ is to be considered an auxiliary field and observables are only made of
the Yang-Mills fields. For such an observable ${\cal O}$,
\beqar
\la {\cal O}\ra &=& \int d\mu (A) \, [d\Phi] \, ~~\Psi^*_{YM} \Psi_{YM}~~
\Psi^*_\Phi \, \Psi_\Phi ~~{\cal O} = \int d\mu (A) \, [d\Phi] \, ~~ \Psi^*_{YM} \Psi_{YM}~
e^{ - 2 W_\Phi}~~{\cal O}
\nonumber\\
&=&  \int d\mu (A) \, ~~\Psi^*_{YM} \Psi_{YM} \, ~~{1\over \det K } \, ~~{\cal O}
\sim \int d\mu (A) \, ~~\Psi^*_{YM} \Psi_{YM} \, ~~{1\over \det ( - D\bD )} \, ~~{\cal O}\nonumber\\
&\approx& \int d \mu (A) \, \Psi^*_{YM} \Psi_{YM} \, \,\exp\left( m \int \! A^{aT} A^{aT} + \cdots\right)\, \, {\cal O}\label{29}
\eeqar
This is equivalent to using
\beq
\Psi_0 \sim \exp \left[ - {1\over 2} \int A^{aT}_i (x) \left[{\sqrt{k^2 + m^2}~  { {- m}} }\right]_{x,y} ~A^{aT}_i (y) ~+\cdots\right]\label{30}
\eeq
where we used the result $\det ( - D \bD ) = \exp ( 2\, c_A \,S_{wzw}(H) )$.
With this result, we have shown that the effective action (\ref{21}) does indeed lead to the wave function we obtained, at least as far as the leading $2J$-term in the exponent of $\Psi_0$.
The procedure clearly admits systematic improvement. As the next step, we can calculate the
${\cal O} (e)$ terms in $\Psi_0$ resulting from the action (\ref{21}) and compare with the 
${\cal O} (e)$ terms of $\Psi_0$ as calculated from the Schr\"odinger equation.
If these do not match, we can improve $\Gamma$ by the addition of a gauge-invariant
monomial with at least three $A$'s (such as $\sim F^3$) to obtain a match. We can continue this procedure to higher orders in $e$, thus using the solution of the Schr\"odinger equation 
to obtain $\Gamma$ in a systematic fashion. This will be considered in a future publication.

As emphasized before, the spinor field which is bosonic must be regarded as an auxiliary field and as a short-hand way of writing a nonlocal term. This way is useful because of the way the expansion scheme works.
We use such a field so that we can get exactly $\det (-D \bD)$ in (\ref{29}); if a scalar (spin zero) field is used, we would get $\det ( - (D \bD + \bD D))$, which does not reproduce $S_{wzw}(H)$
exactly.

\section{Comments, discussion}

Equation (\ref{21}) which gives the leading terms in $\Gamma$ which give the $\Psi_0$
as in (\ref{12}), or (\ref{30}), is the main result of this paper. The rest of this paper will be made of
some comments and discussion about the nature of this $\Gamma$. This is in the nature of a first look at the new directions suggested by $\Gamma$ and which are currently under investigation.

One of the issues which arises in considering a massive gluon field is the following.
Exchange of massive gluons would suggest short range forces or potentials; how can this be compatible with the existence of long range potentials as implied by the area law for the Wilson loop?
This has led to the suggestion that there must be some kind of auxiliary massless fields in the problem \cite{cornwall1}. It is eminently sensible to identify the field $\Phi^a_A$ with the expected massless field.
The crucial $-m/k^2$ term in the kernel in the wave function
(\ref{10}) arises from $\Phi^a_A$; this is also in agreement with this identification.

We now turn to the nature of the mass term. One may think of it, in the context of the effective action, as arising from resummations using a seed mass term, as has been done by a number of authors
\cite{AN, mass}.
The results have slight variations depending on the seed mass term used.
In these cases, it is useful to ask about the nature of threshold singularities \cite{JP2}.
The mass term of \cite{AN} has no singularities at  $p^2 = 0$ for
the one-loop contribution to the gluon propagator, other suggested expressions do.
This suggests that even though we have obtained a mass for the gluon via resummation, 
there are still some massless (likely composite) fields in the problem; these are revealed by considering unitarity cuts of the one-loop contribution. (Strictly speaking, the imaginary part has the wrong sign, corresponding to a magnetic-type instability; this can be interpreted in terms of massless fields with additional magnetic moment interactions.)
So for the mass term to be used in
(\ref{21}), the minimal choice would be the mass term in \cite{AN}.
This does not mean that other choices are to be ruled out; rather, other choices are possible, 
and can be used with some modification of the $\Phi$-terms.

One of the most important results suggested by the effective action (\ref{21}) is the possible existence of ${\mathbb Z}_N$-vortices. For an ordinary scalar field in the adjoint representation coupled to $A$, 
the energy functional for static fields, choosing $A_0 =0$, is of the form
\beq
{\cal E} = \int \left[ {1\over 2} B^2  + 
(D_i \phi)^* (D_i \phi) + \lambda ( \phi^* \phi - (v^2/2) )^2 \right]
\label{51}
\eeq
with $D_i \phi = \del_i - i e A_i \phi$. This admits topological vortex solutions of vortex number $Q$
\cite{vortex} where
\beq
\int d^2x \, F = {2\pi \over e} \, Q
\label{52}
\eeq
Finiteness of energy requires that $D_i \phi$ go to zero at spatial infinity. An ansatz of the form
\beq
\phi = {v \over \sqrt{2}} \, h(r) \, e^{i\theta}, \hskip .2in e A_i = - {\epsilon_{ij} x^j \over r}
\, f(r)
\label{53}
\eeq
where $f$ and $h$ are zero at $r= 0$ and go to $1$ as $r\rightarrow \infty$ will give the single vortex solution. In this case, $A_i$ goes to a pure gauge and $D_i \phi$ vanishes as $r \rightarrow \infty$.
The rate of approach to the asymptotic value is controlled by $ev$ for the vector field and 
$\sqrt{2\lambda}\, v$ for the scalar field. In particular, for $\lambda \ll e^2/2$, the scalar field
is spread out over a large range of $r$.

In our case, for static fields with $A_0 =0$, we have
\beq
\Gamma = 4 \int d^2x\, \left[ ( \bD \Phi_2)^\dagger (\bD \Phi_2) ~+~
(D \Phi_1)^\dagger (D \Phi_1) \right]
\label{54}
\eeq
Vortices are obtained by considering a gauge field of the form
\beq
e A_i = - {\epsilon_{ij} x^j \over r}  f (r) \, Y
\label{55}
\eeq
where $Y$ is the diagonal element of the Lie algebra which exponentiates to the
${\mathbb Z}_N$ elements.
In the fundamental representation, it is the matrix
\beq
Y = diag ( {1\over N} , {1\over N}, \cdots, {1\over N}, -1 + {1\over N} )
\label{56}
\eeq
We see from (\ref{54}) that $\Phi^a_A$ can go to a nonzero constant value at spatial infinity, up to a gauge transformation. An ansatz of the form (\ref{53}) will give finite energy for the
$\Phi$-part of $\Gamma$.
As for the gauge field part, the Yang-Mills action will be as in the scalar field case. The mass term, being gauge-invariant, will also have a rapidly decreasing integrand.
We expect to get  a finite value for the integral.
Thus, using our $\Gamma$, we can get vortices of winding number $Q$, the magnetic flux being
$2 \pi Q /e$. The holonomy at spatial infinity for the
field configurations (\ref{55}) in the fundamental representation then gives an element of
${\mathbb Z}_N$, namely, $\exp(2\pi i /N)$ for elementary vortex. For fields in the adjoint representation, the holonomy will be $1$.

In $\Gamma$, we do not have the analogue of the scalar potential energy. However, consider,
for the sake of the argument, $\Gamma$ with an additional term
\beq
V = \lambda \int d^2x\, \left[( \Phi^a_A)^* \Phi^a_A) - {v^2 \over 2} \right]^2
\label{57}
\eeq
The constant $\lambda$ now controls the profile of the field $\Phi$.
We can see that, as we let $\lambda$ go to zero, the vortices would spread out over an increasing range of $r$. The energy of the vortices will become smaller as well.

Another key difference in our case is that $\Phi^a_A$ is a spinor under Lorentz transformations.
The behavior under Lorentz transformations can be analyzed by introducing a collective coordinate for these via
\beq
\Phi^a_A = g_{AB} \, \Phi^{(0)a}_B
\label{58}
\eeq
where $\Phi^{(0)a}_B$ is a particular solution for the vortex and $g_{AB}$ is a Lorentz matrix depending on the time-variable $\tau$.
Using this in $\Gamma$, we get terms like
\beqar
\Gamma &=&  - \int d\tau\,  \Tr \left[ {\cal I} \, ( g^{-1} \del_0 g \, g^{-1} \del_0 g )\right] 
~+ ~\cdots\nonumber\\
{\cal I}_{BA} &=& \int d^2x\,  \Phi^{(0)a\dagger}_A \Phi^{(0)a}_B
\label{59}
\eeqar
In general, if $\Phi$ approaches a nonzero value at spatial infinity,
${\cal I}_{BA}$ will be divergent. (This is very similar to what happens with the issue of global color for monopoles \cite{globalcolor}.)
Thus, even though vortices can exist, for Lorentz invariance,
we will need net zero vortex number so that the holonomy at spatial infinity is zero, and, correspondingly,  the asymptotic value of
$\Phi$ is zero.
Thus the only allowed configurations are a gas of vortices and antivortices such that the net vortex number is zero. 

What are the physical implications of these vortices? We may expect, in accordance
with the arguments of many authors \cite{Zn}, that these vortices play  a key role in
the screening of the screenable representations (${\mathbb Z}_N$-invariant representations).
It is possible that with a proliferation of vortices the contribution of 
the $\vert \sigma\cdot D \Phi\vert^2$-term to the wave function is altered, may be eliminated; this could lead to a scenario for deconfinement, as the crucial $-m/k^2$-term in
$\Psi_0$ is lost. It would also be interesting to connect this with the gluelump
state analyzed in \cite{AKN}.
The possible existence of the vortices is very suggestive for the issue of screening.
However, as mentioned above, the allowed configurations must have net vortex number equal to zero.
Further, considering that these are also rather spread-out configurations, their importance to physics needs more detailed investigation.

We have outlined a general procedure for analyzing the excited states as well. In applying this to the Yang-Mills theory for glueball states, we must look for gauge-invariant combinations, rather than just products of the fields like $c_k c_l$.
In other words, we must consider shifts of the action of the form
\beq
{\widetilde S}_M = S_M + \int \xi (\vx) \, {d {\cal O}\over dt}
\label{60}
\eeq
where ${\cal O}$ is a gauge-invariant monomial of the fields with zero color charge
(like $F^2$ for the tower of $0^{++}$ glueballs). A corresponding modified version
of (\ref{20p}) can then be obtained. As mentioned after (\ref{20u}), the four-point and higher point
terms can lead to potentials between the constituent $A$'s in ${\cal O}$ which involve
$\la A_i (\vx ) \, A_j (\vy )\ra$.
Since such expectation values
are calculated with the full ground sate wave function (and hence the two-dimensional Yang-Mills action), we can expect terms proportional to a linear potential to appear in the many-particle equations.
While the derivation of these equations is a tedious and difficult task, 
there will be at least one advantage for the gauge fields compared to non-gauge field theories:
The mixing between glueball states is suppressed at large
$N$ \cite{witten}, so single glueball equations should be obtainable in this limit.

It is also  interesting to see how our procedure for the action applies to some of the other approaches using wave functions. For example, Kogan and Kovner have suggested the use of variational wave functions for compact electrodynamics \cite{kogan1}. They have also used a similar strategy, resulting in a wave function which is somewhat different, for QCD in 3+1 dimensions \cite{kogan2}.
Their solution for compact electrodynamics in 2+1 dimensions is \cite{kogan1}
\beqar
\Psi_0 &\sim& \exp \left[ - {1\over 2} \int A^{T}_i (x)~ G (x,y) ~A^{T}_i (y) ~+\cdots\right]\nonumber\\
&\sim&\exp \left[ - {1\over 2} \int A^{T}_i (x) \left[{\sqrt{k^2 + m^2}~ - { {m^2\over \sqrt{k^2+m^2}}} }\right]_{x,y} ~A^{T}_i (y) ~+\cdots\right]\label{61}
\eeqar
where $G(x,y)$ is variationally determined and, in the second line of (\ref{61}), we have used the variational solution they have obtained. The parameter $m$ is essentially arbitrary; it can be related to other parameters of the theory but that formula involves the (arbitrary) value of the upper cut-off on momenta. The kernel $\sqrt{k^2 +m^2} - (m^2 /\sqrt{k^2 + m^2})$ in (\ref{61}) differs from our kernel 
in (\ref{12}) only in the second term, and, indeed, the second term reduces to $-m$ for low momentum modes and agrees with our formula,
except for $m$ being a free parameter. 
Therefore, the effective action for this case can be written as an Abelian version of(\ref{21}).
Instead of the $\Phi$-fields in the adjoint representation, we should have
complex $\Phi_A$ coupling to the electromagnetic field with a charge $e^*$, with
$m = e^{*2}/4\pi$ and $D_\mu \Phi_A = \del_\mu \Phi_A - i e^* A_\mu \,\Phi_A$.
Thus the action is
\beq
\Gamma = \int {1\over 4} F_{\mu\nu} F_{\mu\nu} + S_{m}(A) 
+ (\sigma^\mu D_\mu \Phi_A)^{\dagger} (\sigma^\nu D_\nu \Phi_A)~+~ \cdots
\label{62}
\eeq
For the 3+1 dimensional case, the calculation of the effective action cannot be taken to this stage, because the crucial result that the Dirac determinant leads to an $A^2$-type term (the passage from the second to the third line of (\ref{29})\,) is not obtained.
Nevertheless, it is an interesting case to study, but is beyond the scope of this paper.

Another wave function which is closely related to ours is in the work of Leigh, Minic and Yelnikov
\cite{LMY}. The kernel they have used involves Bessel functions and the explicit calculation of the effective action has proven to be impossible so far.
However, if we restrict attention to the terms quadratic in the currents in their approach as well,
the low and high momentum limits agree with ours and, for all momenta,
the kernel is very close to ours; see the numerical comparison in
\cite{nair-trento1}. Therefore the effective action (\ref{12}) should be a very good approximation for the LMY
wave function as well.

Finally, we can ask whether the procedure for identifying the effective action from the wave function can be applied to other simple systems for which the wave functions are known.
The BCS wave function for superconductivity is an interesting example. \!\!\footnote{I thank the referee for bring this example to my attention.}
In this case, the analysis presented in this paper is not directly applicable,
we need a fermionic version. We hope to take this up in a future publication.

\bigskip

I thank Dimitra Karabali for a critical reading of the manuscript.
This work was supported by U.S.\ National Science
Foundation grant PHY-0855515
and by a PSC-CUNY award. 

%%%%%%%%%%%%%%%%
%%%%%%%%%%%%%%%%

%%%%%%%%%%%%%%%%%%%%%%%%%%%%%%%%%%%%%%%%%%%%%%%%
%%%%%%%%%%%%%%%%%%%%%%%%%%%%%%%%%%%%%%%%%%%%%%%%

\begin{thebibliography}{99}

\bibitem{KN1} D. Karabali and V.P. Nair,
\NP~ {\bf B464}, 135 (1996);
\PL~ {\bf B379}, 141 (1996).

\bibitem{KKN1} D. Karabali, Chanju Kim and V.P. Nair, \NP~ {\bf B524}, 661 (1998);

\bibitem{nair-trento1}
For a recent review, see V.P. Nair,
Invited talk at the {\it Workshop on QCD GreenÕs Functions, Confinement and Phenomenology, QCD- TNT}, September 2009, Trento, Italy; published in
{\it Proceedings of Science, POS(QCD-TNT09) 030}, 
\verb+http://pos.sissa.it//archive/conferences/087/030/QCD-TNT09_030.pdf+


\bibitem{KKN2} D. Karabali, Chanju Kim and V.P. Nair,
\PL~ {\bf B434}, 103 (1998).

\bibitem{KNY} D. Karabali, V.P. Nair and A. Yelnikov, \NP~ {\bf B824}, 387 (2010).

\bibitem{KNrobust} D. Karabali and V.P. Nair, 
Phys. Rev. {\bf D77}, 025014 (2008)

\bibitem{cornwall1} J.M. Cornwall, \PR~{\bf D76}, 025012 (2007).

\bibitem{AN} V.P. Nair,  \PL~ {\bf 352 B}, 117 (1995);
G. Alexanian and V.P. Nair,  \PL~{\bf 352 B},  435 (1995).


\bibitem{mass} J.M. Cornwall, Phys. Rev. {\bf D10}, 500 (1974); 
{\it ibid.} {\bf D26}, 1453 (1982); {\it ibid.} {\bf D57}, 3694 (1998);
J. M. Cornwall, W. S. Hou and J. E. King, \PL~{\bf B 153}, 173 (1985);
J. M. Cornwall and B. Yan, \PR~{\bf D 53}, 4638 (1996).
W. Buchmuller and O. Philipsen, \NP~{\bf B443} (1995) 47;
O. Philipsen, in {\it TFT-98: Thermal Field Theories and
their Applications}, U. Heinz (ed.), hep-ph/9811469;
R. Jackiw and S-Y. Pi, \PL~{\bf B 403}, 297 )1997).

\bibitem{JP2} R. Jackiw and S.Y. Pi, \PL~ {\bf B368}, 131(1996).

\bibitem{vortex} A.A. Abrikosov, Sov. Phys. JETP 5, 1174
(1957); H.B. Nielsen and P. Olesen, Nucl. Phys. B61, 45 (1973);
see also, A. Jaffe and C. Taubes, {\it Vortices and Monopoles: The Structure of Static Gauge Theories}, Progress in Physics 2, Birkhauser (1980).


\bibitem{Zn} There is a large body of  literature on this topic, going back to
"t Hooft's emphasis of the role ${\mathbb Z}_N$ symmetry in
G. 't Hooft, \NP~{\bf B153}, 141 (1979). For a rceent survey, see,
J. Greensite, Eur. Phys. J. ST. {\bf 140}, 1 (2007).

\bibitem{globalcolor} P. Nelson and A. Manohar, \PRL~{\bf 50}, 943 (1983);
A.P. Balachandran {\it et al}, \PRL~{\bf 50}, 1553 (1983);
A. Abouelsaood, \NP~{\bf B 226}, 309 (1983);
P. Nelson and S. Coleman, \NP~ {\bf B 237}, 1 (1984).

\bibitem{AKN} A. Agarwal, D. Karabali and V.P. Nair,
Nucl. Phys. {\bf B790}, 216 (2008).

\bibitem{witten} E. Witten, \NP~{\bf B 160}, 57 (1979).

\bibitem{kogan1} I.I. Kogan and A. Kovner, \PR ~{\bf D51}, 1948 (1995).

\bibitem{kogan2}I.I. Kogan and A. Kovner, \PR~ {\bf D52}, 3719 (1995).

\bibitem{LMY} R.G. Leigh, D. Minic and A. Yelnikov, \PRL ~{\bf 96}, 222001 (2006);
\PR~{\bf D76}, 065018 (2007).

\end{thebibliography}
\end{document}